\documentclass[a4paper,
              ]{jacow}
\makeatletter%
	\ifboolexpr{bool{xetex}}
	 {\renewcommand{\Gin@extensions}{.pdf,%
	                    .png,.jpg,.bmp,.pict,.tif,.psd,.mac,.sga,.tga,.gif,%
	                    .eps,.ps,%
	                    }}{}
\makeatother

\ifboolexpr{bool{xetex} or bool{luatex}} 
 {}                                      
 {\usepackage[utf8]{inputenc}}           

\usepackage[USenglish]{babel}			 

\usepackage[final]{pdfpages}
\usepackage{multirow}
\usepackage{ragged2e}

\ifboolexpr{bool{jacowbiblatex}}%
 {%
  \addbibresource{jacow-test.bib}
  \addbibresource{biblatex-examples.bib}
 }{}
\def\unsty{\,\hbox}

\begin{document}

\title{IMPEDANCE REDUCTION OF THE HIGH-FREQUENCY CAVITIES\protect\\IN THE CERN PS BY
MULTI-HARMONIC FEEDBACK}

\author{F. Bertin, Y. Brischetto, H. Damerau, A. Jibar, D. Perrelet, CERN, Geneva, Switzerland}
	
\maketitle

\begin{abstract}
In the framework of the "LHC Injectors Upgrade" (LIU) project, the CERN Proton Synchrotron (PS) is being prepared as a pre-injector for the Large Hadron Collider at high luminosity (HL-LHC). RF systems at 20 MHz, 40 MHz and 80 MHz are required for longitudinal beam manipulations to provide 25 ns bunch spacing and short bunches at transfer to the Super Proton Synchrotron (SPS). Beam induced voltage in these cavities causes transient beam-loading and uncontrolled longitudinal emittance blow-up. To reduce the impedance beyond the reach of the wide-band feedback around the power amplifier, a new multi-harmonic feedback has been developed and commissioned. Based on narrow-band filters treating each revolution frequency harmonic separately, the feedback phase at these harmonics can be dynamically adapted during acceleration, taking into account the phase of the cavity transfer function. The beneficial effect on longitudinal beam quality of LHC-type beams is shown.
\end{abstract}

\section{INTRODUCTION}

The CERN Proton Synchrotron (PS) is being prepared for its role as pre-injector to the High-Luminosity LHC (HL-LHC). In particular, the intensity per bunch doubles at constant bunch length and longitudinal emittance from $1.3\cdot10^{11}$ particles per bunch (p/b) in trains of 72 bunches to $2.6\cdot10^{11}\unsty{p/b}$, effectively doubling the longitudinal density. Dipole and quadrupole coupled-bunch instabilities pose major limitations to longitudinal stability and only the dipole oscillations are controlled by a dedicated wide-band feedback system. Additionally, the longitudinal beam quality is compromised by an uncontrolled longitudinal emittance growth along the batch, as well as by transient beam loading during the RF manipulations resulting in an increasing bunch-by-bunch parameter spread, in particular in the intensity range above about $2.1\cdot10^{11}\unsty{p/b}$.

For the production of LHC-type beams in the PS as many as 25~RF cavities in the frequency range from about $400\unsty{kHz}$ up to $200\unsty{MHz}$ are installed. The eleven tunable $2.8\unsty{MHz}$ to $10\unsty{MHz}$ ferrite loaded cavities are already equipped with direct~\cite{garoby1989} and upgraded 1-turn delay feedback systems~\cite{blas1991,perrelet2013}. The $20\unsty{MHz}$ cavities are only damped by direct feedback systems, but their gaps remain short-circuited during the majority of the cycle. However, the impedances of the two $40\unsty{MHz}$~\cite{garoby1997} and three $80\unsty{MHz}$~\cite{jensen1998} cavities are reduced only by direct feedback systems, and the gaps of these vacuum resonators must remain open permanently. For the bunch rotation to shorten the proton bunches prior to extraction two cavities at $80\unsty{MHz}$, with a voltage of $300\unsty{kV}$, are active. For parallel operation of proton and ion beams, the third $80\unsty{MHz}$ cavity is operated at slightly lower resonance frequency to adiabatically shorten the lead ($^{208}$Pb$^{54+}$) ion bunches. It became clear that the residual impedance of all three cavities affects the quality of the proton beam~\cite{damerau2017a}.

\begin{figure}[!tbh]
	\centering
	\includegraphics*[width=0.6\columnwidth,bb=145 249 445 592]{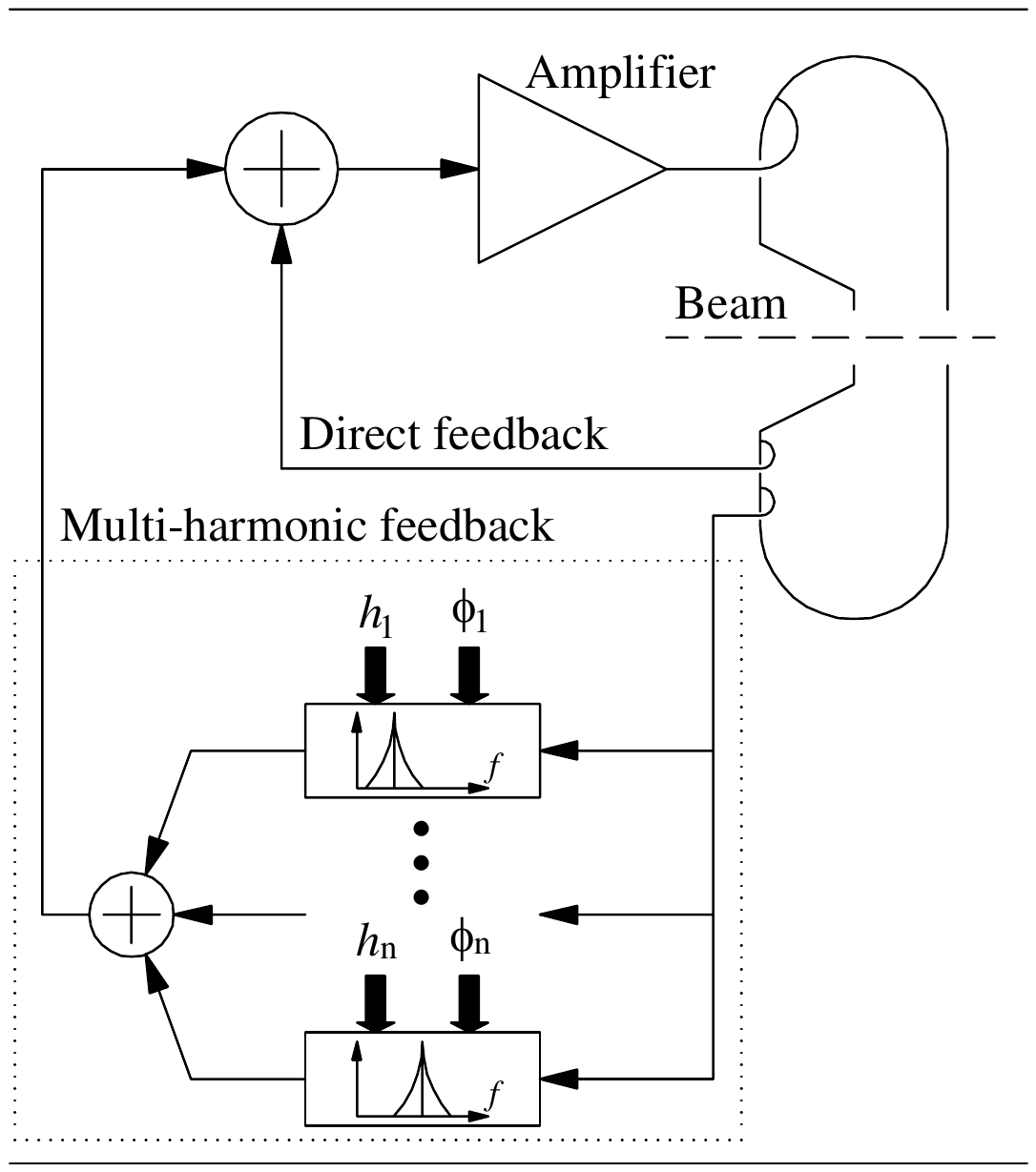}	
	\caption{Simplified overview diagram of a $40\unsty{MHz}$ or $80\unsty{MHz}$ RF system with direct and new multi-harmonic feedback. The automatic voltage control (AVC) loop is not shown.}
	\label{figSchematicOverviewMultiHarmonicFeedback}
\end{figure}

A multi-harmonic feedback system (Fig.~\ref{figSchematicOverviewMultiHarmonicFeedback}) has therefore been designed which treats every harmonic of the revolution frequency by a separate signal processing~\cite{boussard1988}. 
This does not only allow to perfectly match the phase of the feedback channels to the non-linear phase of the cavity resonator with frequency, but also to dynamically adjust the phase of each channel during acceleration. 

\section{MULTI-HARMONIC FEEDBACK}

The maximum gain, $G_\mathrm{max}$ of conventional direct feedback systems~\cite{garoby1989} is limited by a phase rotation of $45^0$ when the open loop gain drops to unity~\cite{boussard1985,baudrenghien2000}, respectively
\begin{equation*}
G_\mathrm{max} = \frac{\pi}{2} \frac{1}{R/Q} \frac{1}{\omega_0} \frac{1}{\tau} = \frac{\pi}{2} \frac{1}{R} \frac{1}{\Delta \omega_{\mathrm{-3dB}}} \frac{1}{\tau} \, ,
\end{equation*}
where $R$ is the shunt impedance of the bare cavity, $Q=\Delta\omega_0/\omega_{\mathrm{-3dB}}$ its unloaded quality factor and $\tau$ the overall loop delay of the feedback. This implicitly assumes that the cavity itself is the only bandwidth limiting element in the loop. Significantly higher feedback gain can be achieved by artificially reducing the bandwidth. Keeping in mind that the longitudinal spectrum of the beam is located at and nearby the harmonics of the revolution frequency, the cavity impedance only needs to be reduced in these limited frequency ranges. 

Although the $40\unsty{MHz}$ and $80\unsty{MHz}$ RF systems are equipped with direct feedback systems attenuating the beam coupling impedance by more than $40\unsty{dB}$~\cite{benedikt2000}, perturbations of the high-intensity LHC-type beam are observed at the flat-top due to the impedance of the $80\unsty{MHz}$ cavities~\cite{damerau2017a}.

A conventional 1-turn delay feedback~\cite{boussard1985, blas1991} would be difficult to realize for the high frequency RF systems. The cavities are at fixed resonance frequency while the revolution frequency of the beam changes by about $5\unsty{\%}$ ($E_\mathrm{kin}=2\unsty{GeV}$ at injection) during acceleration, sweeping several beam harmonics through the resonance curve of the beam. Additionally, due to the narrow bandwidth of the cavities, the non-linear phase of the resonator transfer function would again limit the gain. The limitations are overcome by treating each harmonic of the revolution frequency by a separate narrow-band feedback, adjusted to compensate the phase of the cavity transfer function.

The resonance curves of the $40\unsty{MHz}$ and $80\unsty{MHz}$ cavities, reduced by the direct feedback, cover about three revolution frequency harmonics on each side of the resonance frequency at $84 f_\mathrm{rev}$ ($f_\mathrm{res} = 40.053\unsty{MHz}$) and $168 f_\mathrm{rev}$ ($f_\mathrm{res} = 80.106\unsty{MHz}$). Hence at constant revolution frequency seven narrow-band feedback chains would be sufficient. However, the revolution frequency of the proton beams in the PS sweeps from $452\unsty{kHz}$ ($E_\mathrm{kin} = 2\unsty{GeV}$) to $477\unsty{kHz}$ ($E_\mathrm{tot} = 26\unsty{GeV}$). Before the upgrade of the injection energy during the Long Shutdown 2 (LS2) this frequency swing was even twice larger. At injection energy, the harmonic at $180 f_\mathrm{rev}$ touches the upper end of the $f_\mathrm{res} \pm 3 f_\mathrm{rev}$ frequency range of the $80\unsty{MHz}$ cavity. The lowest harmonic to be treated by the multi-harmonic feedback is $h=168 -3 = 165$. Ideally, a total of 16 signal processing chains would thus cover all relevant harmonics for the $80\unsty{MHz}$ cavities and somewhat fewer harmonics for the cavities at $40\unsty{MHz}$. However, the long bunches at injection have little spectral content around $40\unsty{MHz}$ or above. These harmonics only become relevant with the RF manipulations at $E_\mathrm{kin} = 2.5\unsty{GeV}$, moving to a fundamental harmonic of $h=21$ and shorter bunches. This reduces the number of spectral components to be treated to $13$.

\section{IMPLEMENTATION}

The multi-harmonic feedback uses a hardware platform originally developed for the 1-turn turn delay feedbacks of the main accelerating cavities~\cite{perrelet2013}. It features four ADC/DAC channels and a Stratix II FPGA. To avoid saturation of the ADC when the cavity is pulsing, a narrow-band surface accoustic wave (SAW) notch filter removes the central RF component from the spectrum of the cavity return signal when the cavity is driven by its power amplifier.

\subsection{Single-harmonic signal processing}
\begin{figure}[!tbh]
	\centering
	\includegraphics*[width=0.9\columnwidth,bb=100 132 495 707]{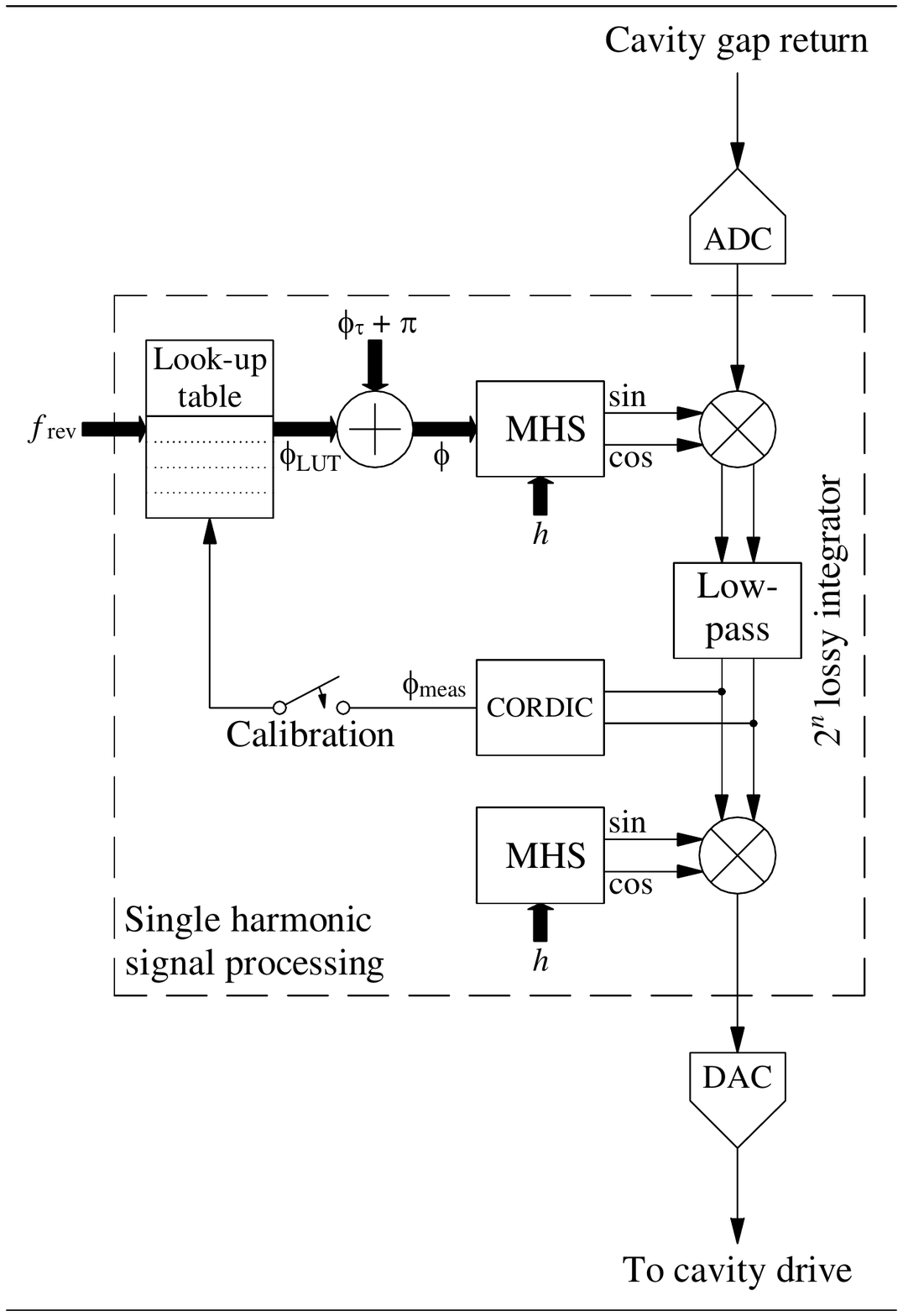}	
	\caption{Digital signal processing for one harmonic of the revolution frequency. The ADC and DAC are common to all signal processing chains.}
	\label{SingleHarmonicSignalProcessing}
\end{figure}

The signal processing to treat one revolution frequency harmonic is sketched in Fig.~\ref{SingleHarmonicSignalProcessing}.
The cavity return signal is directly digitized once and transmitted to the FPGA which implements multiple signal processing chains simultaneously. The relevant parts of the FPGA are clocked at $h_\mathrm{clk} = 256$ which means that the system runs in baseband at $h=84$~($40\unsty{MHz}$) and under-samples at $h=168$~($80\unsty{MHz}$). Keeping in mind that the digital filters are narrow-band, operating in the second Nyquist zone causes no additional complication other than an analogue filter to remove components in the baseband at $h=256-168=88$ and higher amplification to adapt the signal levels.

The gap return signal from the cavity is first digitally down-converted to base-band I/Q data. A beam synchronous multi-harmonic source (MHS) serves as a numerical oscillator~\cite{damerau2017b}. All parameters, in particular its harmonic number and phase are programmable and a virtual delay can be added by introducing a frequency dependent phase, $\phi_\tau = \omega \tau$~(Fig.~\ref{SingleHarmonicSignalProcessing}). The phase of the down-converting MHS defines the overall phase of the signal processing chain. The baseband I/Q data streams are then low-pass filtered using a lossy integrator filter to keep only the spectral content around the selected revolution frequency harmonic. It allows to change filter characteristics easily while consuming only moderate logic resources in the FPGA. The filtered data streams are transposed back to the selected harmonic by a simplified MHS at fixed phase and reconverted to analogue. Conceptually, the signal processing chain represents a narrow passband filter with programmable phase, which tracks a selected revolution frequency harmonic.

Special care must be taken to set the filter phase, $\phi$. Some of the revolution harmonics actually sweep through the resonance curve of the cavity, e.g. the signal processing chain at $h=h_\mathrm{RF}+2=170$ is well below the resonance frequency of the $80\unsty{MHz}$ cavity at low beam energy, but well above it at the flat-top. This implies that a fixed delay and phase are not sufficient and that the phase curve of the cavity transfer function must be compensated. A look-up table~(Fig.~\ref{SingleHarmonicSignalProcessing}) sets the filter phase according to the measured revolution frequency of the beam.

To fill the look-up table automatically, an integrated calibration system is provided. In open loop the phase of the cavity return is extracted from the I/Q baseband signal through a co-ordinate rotation digital computer~(CORDIC) and written into the look-up table during a revolution frequency sweep without beam. This provides the necessary phase information to compensate the phase of the cavity transfer function which, with an additional offset of $\pi$, is applied during feedback operation. Closing the feedback loop around the cavity, each signal processing chain generates a symmetric notch in the cavity transfer function at a given revolution frequency harmonic.

\subsection{Closed-loop transfer function}

\begin{figure}[!tbh]
	\centering
	\includegraphics*[width=0.9\columnwidth]{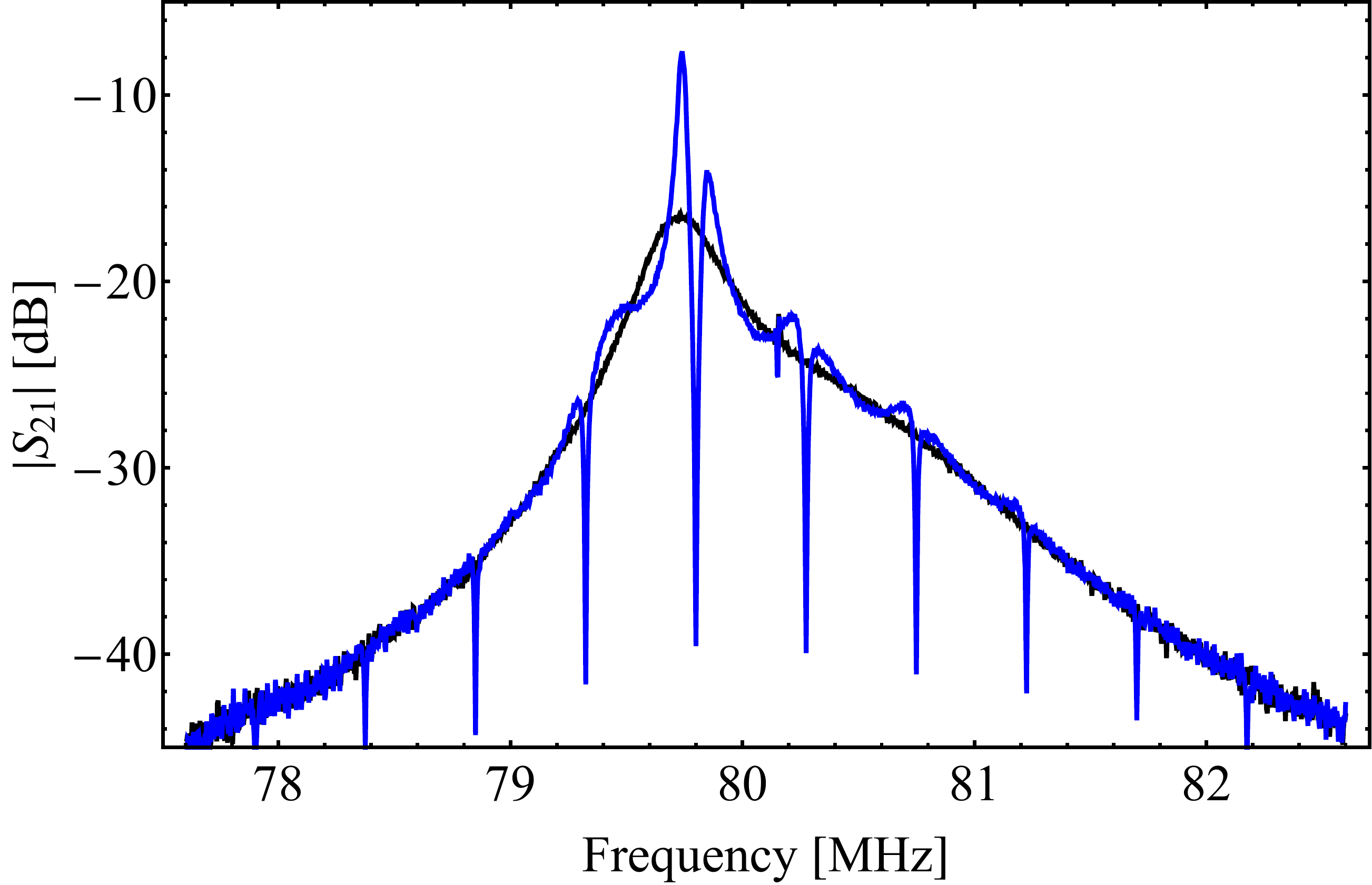}	
	\caption{Transfer function of an $80\unsty{MHz}$ system with~(blue) and without multi-harmonic feedback~(black, direct feedback only). The resonance frequency of the cavity, slightly below $80\unsty{MHz}$, was set for ion operation. The spurious line between $f_\mathrm{RF}$ and $f_\mathrm{RF} + f_\mathrm{rev}$ is caused by leakage from the LLRF.}
	\label{OpenClosedLoopTransferFunction80MHz}
\end{figure}

A comparison of the transfer function of an $80\unsty{MHz}$ cavity with and without the multi-harmonic feedback is shown in Fig.~\ref{OpenClosedLoopTransferFunction80MHz}.
The slight asymmetry of the central notch is caused by the phase sensitivity at high feedback gain. However, even the notches far away from the central resonance frequency stay symmetric at all times, which would not be the case for a classical 1-turn delay feedback.

\begin{figure}[!tbh]
	\centering
	\includegraphics*[width=0.9\columnwidth]{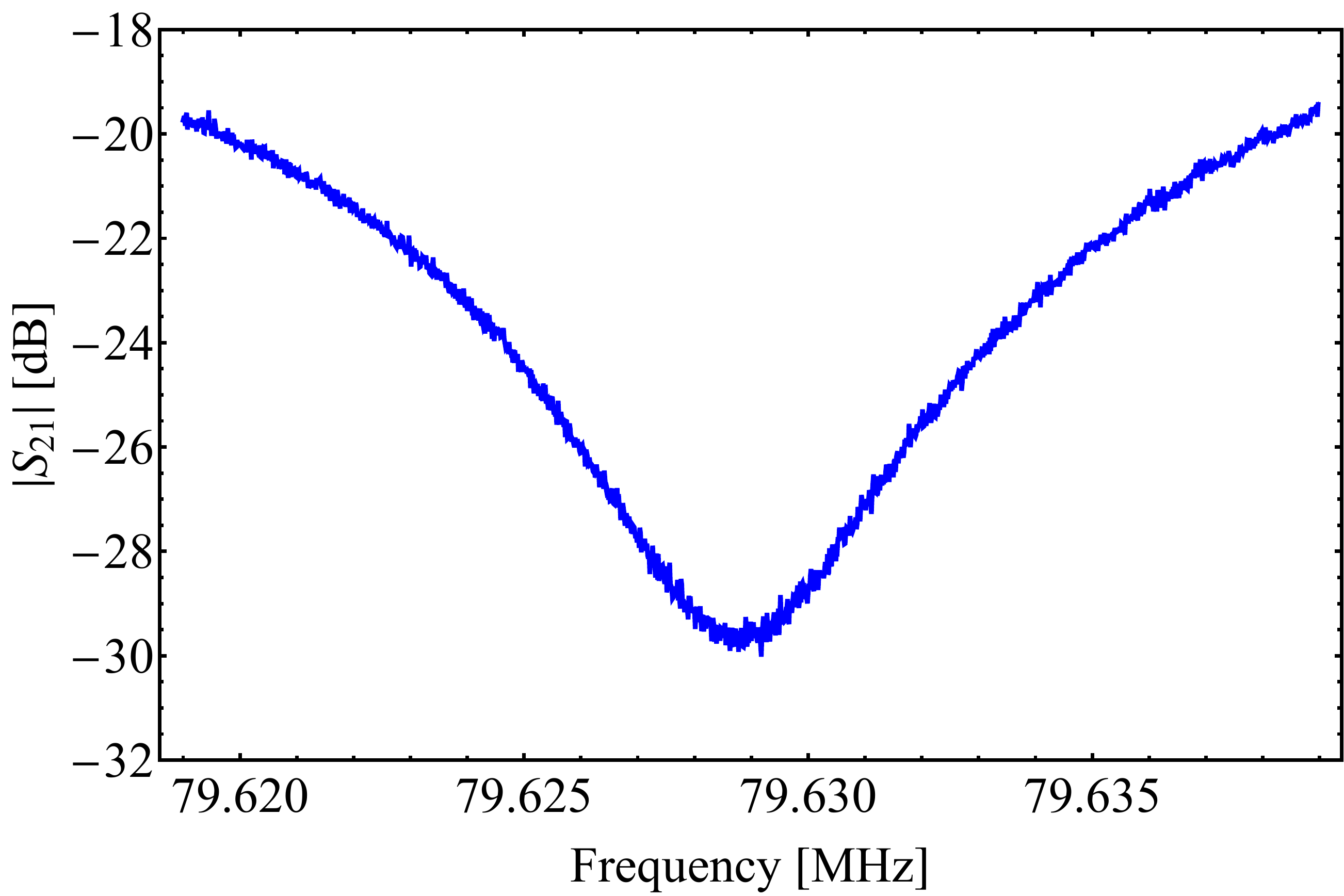}	
	\caption{Transfer function with the multi-harmonic feedback zoomed around a harmonic of the revolution frequency.}
	\label{OpenClosedLoopTransferFunction80MHzSingleNotch}
\end{figure}

Each of the notches is only a few kilohertz wide~(Fig.~\ref{OpenClosedLoopTransferFunction80MHzSingleNotch}), sufficient to cover the corresponding revolution frequency harmonic and its synchrotron frequency sidebands.

\subsection{Results with beam}

\begin{figure}[!tbh]
	\centering
	\includegraphics*[width=0.85\columnwidth]{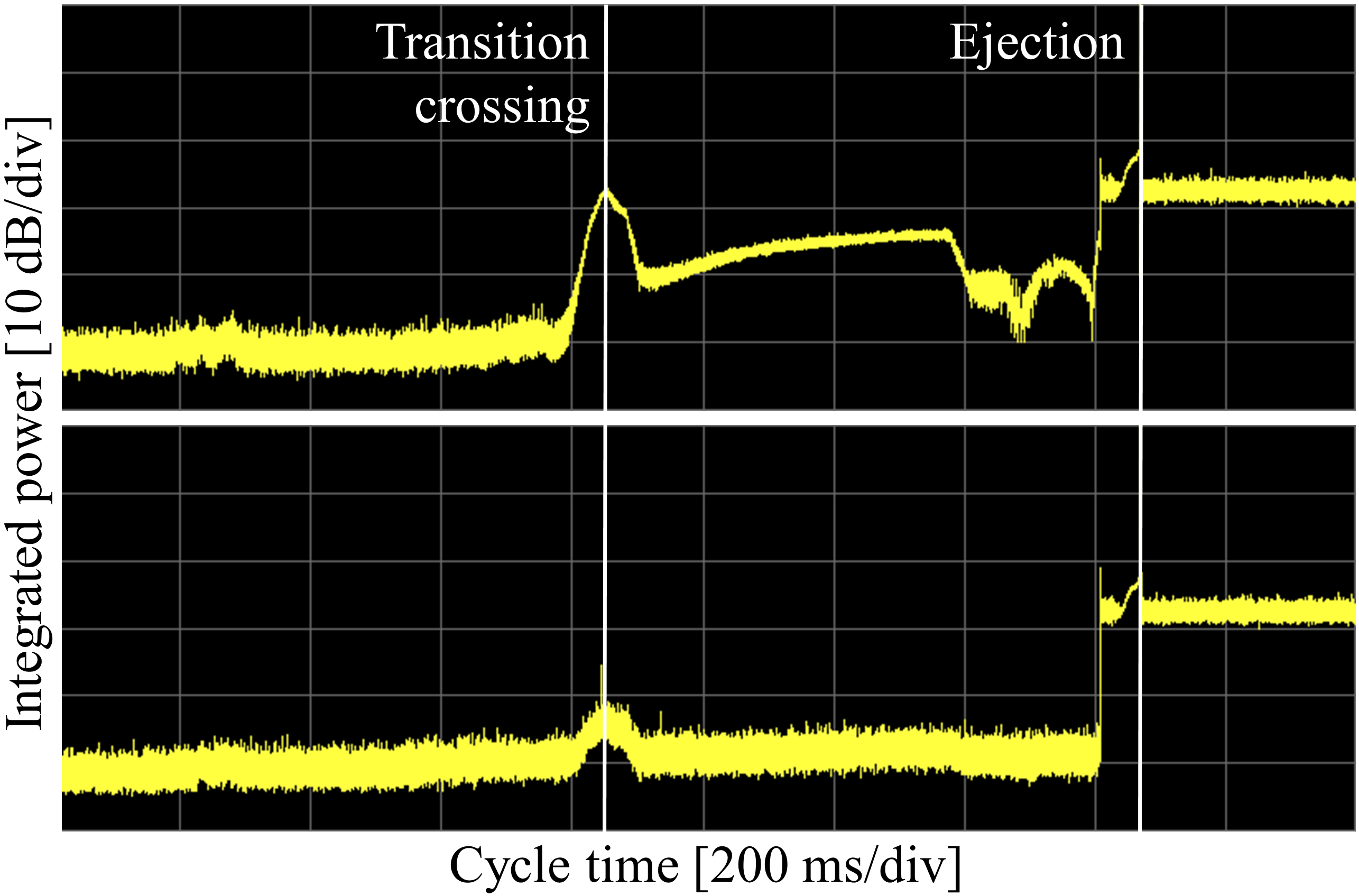}	
	\caption{Beam induced power integrated over a frequency range of $40.05\unsty{MHz} \pm 3\unsty{MHz}$, measured at a probe pick-up of a $40\unsty{MHz}$ cavity during the acceleration of an LHC-type multi-bunch beam without (top) and with the multi-harmonic feedback (bottom). During the last $70\unsty{ms}$ before ejection the power is due to the amplifier driving the cavity at $h=84$.}
	\label{BeamInducedPowerFBOffOn}
\end{figure}

Figure~\ref{BeamInducedPowerFBOffOn} illustrates the measured beam induced power in a $40\unsty{MHz}$ cavity during the acceleration of an LHC-type multi-bunch beam without and with the multi-harmonic feedback.
The power increases around transition crossing when the bunches are shortest, and again at the flat-top when the harmonic number changes from $h=21$ via $42$ to $84$.

\begin{figure}[!tbh]
	\centering
	\includegraphics*[width=0.9\columnwidth]{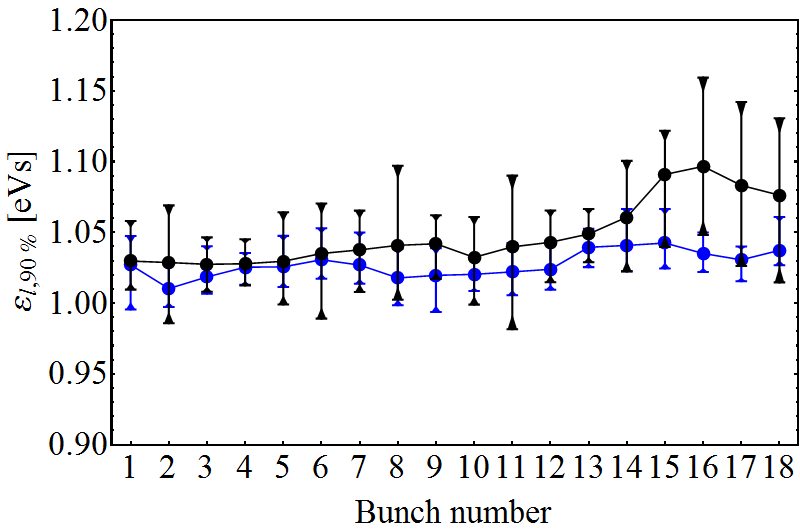}	
	\caption{Longitudinal emittance along a batch of 18~bunches in $h=21$ measured by tomographic reconstruction~\cite{hancock1998} at the arrival of the flat-top without~(black) and with~(blue) the multi-harmonic feedbacks. The intensity corresponds to $2.6\cdot10^{11}\mathrm{p/b}$ at extraction.}
	\label{LongitudinalEmittanceFlatTopMultipleFeedbacksOnOff}
\end{figure}

The beneficial effect of the multi-harmonic feedbacks on the longitudinal beam quality becomes clear from Fig.~\ref{LongitudinalEmittanceFlatTopMultipleFeedbacksOnOff}. 
With the direct feedback only the longitudinal emittance of the last bunches is blown up unacceptably at the arrival of the flat-top. This blow-up is entirely removed with the multi-harmonic feedbacks enabled for all active $40\unsty{MHz}$ and $80\unsty{MHz}$ cavities.

\section{CONCLUSIONS}

A multi-harmonic feedback for the high frequency cavities in the PS has been developed and commissioned with beam on the $40\unsty{MHz}$ and $80\unsty{MHz}$ cavities. The feedback is based on the individual treatment of the relevant revolution frequency harmonics. The phase of the feedback filters is dynamically adjusted with the revolution frequency to compensate for the non-linear phase of the cavity transfer function. This is particularly important to achieve high gain for the harmonics which sweep entirely through the cavity resonance during acceleration. A feedback gain of about $20\unsty{dB}$ is measured which significantly reduces the beam induced voltage. In combination with the coupled-bunch feedback, the multi-harmonic feedback allows to reliably reach the bunch intensity of $2.6\cdot10^{11}\mathrm{p/b}$ in trains of 72~bunches required for the operation of the PS for the HL-LHC. The firmware of the feedback is now being extended to a cavity controller, incorporating also the beam synchronous RF source and the voltage control loop.

\section{ACKNOWLEDGEMENTS}

The authors are grateful to the PSB and PS operations teams for their support of the high-intensity beam tests.

	\AtEndDocument{\par\leavevmode}

\end{document}